\documentclass[aps,prd,preprint,eqsecnum,nofootinbib]{revtex4-1}
\usepackage[english]{babel}
\usepackage[utf8]{inputenc}
\usepackage{amsmath}
\usepackage{amssymb}
\usepackage{graphicx}
\usepackage{verbatim}

\def\Z{\mathbb Z}
\def\C{\mathbb C}
\def\CP{\mathbb{CP}}

\def\mc{\mathcal}
\def\nn{\nonumber}
\def\pd#1#2{\dfrac{\partial#1}{\partial#2}}
\def\spaa#1#2{\langle#1\,#2\rangle}

\def\avg#1{\langle#1\rangle}
\def\Res{\operatorname{Res}}
\def\CHY{{\operatorname{CHY}}}
\def\vol{\operatorname{vol}}
\def\SL{\operatorname{SL}}
\def\Pf{\operatorname{Pf}}

\def\tree{{\operatorname{tree}}}

\def\Lra{\,\Longrightarrow\,}

\newtheorem{thm}{Theorem}

\newtheorem{example}{Example}
\newtheorem{algorithm}{Algorithm}

\begin{document}
\date{\today}
\author{Mads S{\o}gaard}
\affiliation{
SLAC National Accelerator Laboratory, Stanford University, \\
2575 Sand Hill Road, Menlo Park, CA 94025, USA
}
\author{Yang Zhang}
\affiliation{
Institute for Theoretical Physics, ETH Z{\"u}rich, \\
Wolfgang-Pauli-Stra{\ss}e 27, CH-8093 Z{\"u}rich, Switzerland
}

\title{Scattering Equations and Global Duality of Residues}

\begin{abstract}
We examine the polynomial form of the scattering equations by means of
computational algebraic geometry. The scattering equations are the backbone of
the Cachazo-He-Yuan (CHY) representation of the S-matrix. We explain how the
Bezoutian matrix facilitates the calculation of amplitudes in the CHY formalism,
without explicitly solving the scattering equations or summing over the
individual residues. Since for $n$-particle scattering, the size of the
Bezoutian matrix grows only as $(n-3)\times(n-3)$, our algorithm is very
efficient for analytic and numeric amplitude computations.
\end{abstract}

\maketitle

\section{Introduction and Review}
Cachazo, He and Yuan (CHY) recently proposed an intriguing representation of
massless tree-level scattering amplitudes in any space-time dimensions in terms
of a multidimensional complex contour integral of a certain rational function
over an auxiliary coordinate space
\cite{Cachazo:2013gna,Cachazo:2013hca,Cachazo:2013iea,Cachazo:2014nsa,Cachazo:2014xea}.
This approach is an alternative to the traditional program based on Feynman
diagrams. The contour $\mc O$ in question is restricted to enclose the
simultaneous solutions of a set of algebraic constraints referred to as the 
{\it scattering equations}. Let $n$ denote the number of external particles. The
auxiliary coordinate space consists of $n$ puncture points $z_i\in\CP^1$ on the
Riemann sphere, and the scattering equations take the form,
\begin{align}
f_a(z,k) = \sum_{b\neq a}
\frac{s_{ab}}{z_a-z_b} = 0\;,
\quad s_{ab}\equiv(k_a+k_b)^2\;,
\quad a\in A = \{1,2,\dots,n\}\;.
\label{EQ_CHY_EQS}
\end{align}

For the benefit of the reader, let us recall a few basic properties previously
reported elsewhere \cite{Cachazo:2013gna,Cachazo:2013hca}. It can be shown that
eqs.~\eqref{EQ_CHY_EQS} are invariant under $\SL(2,\C)$ transformations,
\begin{align}
z_a\mapsto\zeta_a = \frac{\alpha z_a+\beta}{\gamma z_a+\delta}\;, 
\quad a\in A\;,
\end{align}
by virtue of momentum conservation. This implies that only $(n-3)$ of the
scattering equations are independent. The $\SL(2,\C)$ invariance allows us to
specify three arbitrary coordinates $z_r$, $z_s$ and $z_t$, say, $z_1\to\infty$,
$z_2$ fixed and $z_n\to0$. An important observation is that the number of
solutions to the scattering equations grows factorially as $(n-3)!$.

Let us return to the construction of tree amplitudes within the CHY formalism.
The rational integrand consists of a universal part $d\Omega_\CHY$ and a purely
theory-dependent factor denoted $\mc I$. The universal part is constructed from
the $\SL(2,\C)$ invariant integration measure and the rational functions
$f_a(z,k)$. It is responsible for localizing the integrand onto the joint
solutions of the scattering equations upon integration. The precise form of 
$\mc I$ has recently been explored for a large variety of quantum field theories
including $\varphi^3$-theory, Yang-Mills, Einstein gravity and
Dirac-Born-Infeld \cite{Cachazo:2013iea,Cachazo:2014nsa,Cachazo:2014xea}. We
write schematically,
\begin{align}
\mc A_n^\tree = \oint_{\mc O}d\Omega_\CHY\mc I(z,k)\;,
\label{EQ_CHY_I}
\end{align}
where
\begin{align}
d\Omega_\CHY\equiv
\frac{d^nz}{\vol(\SL(2,\C))}
{\prod_a}'\frac{1}{f_a(z,k)} = 
\prod_{a\in A\backslash\{r,s,t\}}dz_a
(z_{rs}z_{st}z_{tr})
(z_{ij}z_{jk}z_{ki})
\prod_{a\in A\backslash\{i,j,k\}}\frac{1}{f_a(z,k)}\;.
\label{EQ_CHY_MEASURE}
\end{align}
Note that the scattering constraints are imposed in a permutation invariant
manner as the measure is independent of both $\{i,j,k\}$ and $\{r,s,t\}$
\cite{Cachazo:2013hca}. Throughout the paper $z_{ab}\equiv z_a-z_b$.

The natural question to address is how to actually calculate the amplitudes in
this formalism. Actually it is in principle very straightforward to carry out
any CHY integral of the form \eqref{EQ_CHY_I} without specifying the form of the
theory-dependent part of the integrand. In fact, the CHY integral simply reduces
to the sum of the $(n-3)!$ nondegenerate multivariate residues evaluated at the
simultaneous zeros $\mc S$ of the denominator factors in
eq.~\eqref{EQ_CHY_MEASURE},
\begin{align}
\mc A_n^\tree = \sum_{z^*\in\:\!\mc S}
\mc J^{-1}(z,k)(z_{ij}z_{jk}z_{ki})
(z_{rs}z_{st}z_{tr})
\mc I(z,k)\big|_{z=z^*}\;.
\label{EQ_TOTAL_SUM}
\end{align}
The Jacobian associated with the individual residues is
\begin{align}
\mc J(z,k) = 
\det_{\substack{
a\in A\backslash\{i,j,k\} \\
b\in A\backslash\{r,s,t\}}}\!
\bigg(\pd{f_a}{z_b}\bigg)\;.
\end{align}
There exists a closed-form expression for this determinant
\cite{Cachazo:2013hca}, but it is not particularly illuminating for our
purposes. Although the CHY formula \eqref{EQ_CHY_MEASURE} is extremely compact,
it suffers from a practical limitation. Indeed, the instruction to sum over the
$(n-3)!$ residues makes this formalism intractable already at relatively low
multiplicities. The major problem is that the solutions can be very complicated
and are inevitably irrational beyond five external particles
\cite{Weinzierl:2014vwa}. Analytic expressions of the solutions are in general
not attainable. Nevertheless, the final result computed from the $(n-3)!$
multivariate residues is always a simple rational function. Our motivation is to
employ the {\it Bezoutian matrix} method from computational algebraic geometry
to directly evaluate the sum of residues \eqref{EQ_TOTAL_SUM} {\it without
solving the scattering equations explicitly.} (The Bezoutian matrix method has
previously proven valuable in multiloop generalized unitarity cuts
\cite{Sogaard:2014oka}.)

The scattering equations and the CHY formalism have received extensive attention
in the literature recently. Here we attempt to provide a brief overview of the
most important developments. A proof of the CHY representation for
$\varphi^3$-theory and Yang-Mills based on Britto-Cachazo-Feng-Witten (BCFW)
recursion relations \cite{Britto:2004ap,Britto:2005fq} and a generalization of
the scattering equations to massive particles were presented in
ref.~\cite{Dolan:2013isa}. Ref.~\cite{Weinzierl:2014ava} discussed scattering
equations and fermions whereas an equivalent polynomial form of the scattering
equations was derived in ref.~\cite{Dolan:2014ega}. Interestingly, new rules and
techniques for evaluating amplitudes from CHY representations have been
established, see papers by Cachazo and Gomez \cite{Cachazo:2015nwa} and
Baadsgaard, Bjerrum-Bohr, Bourjaily and Damgaard \cite{Baadsgaard:2015voa}.
Moreover, the authors of ref.~\cite{Baadsgaard:2015ifa} uncovered a link between
CHY integrals and individual Feynman diagrams. Additional novel insight was
achieved through string theory
\cite{Mason:2013sva,Berkovits:2013xba,Gomez:2013wza,
Ohmori:2015sha,Casali:2015vta,Bjerrum-Bohr:2014qwa,Geyer:2014fka,
Lipstein:2015rxa,Lipstein:2015vxa}. Despite the fact that the present paper is
concentrated on tree-level amplitudes, we should mention that, very recently,
Geyer, Mason, Monteiro and Tourkine \cite{Geyer:2015bja} generalized the
scattering equations at one loop \cite{Adamo:2013tsa} and suggested how to
extend the result to arbitrary loop order. This analysis was further generalized
by Baadsgaard, Bjerrum-Bohr, Bourjaily, Damgaard and Feng
\cite{Baadsgaard:2015hia} for one-loop amplitudes in $\varphi^3$-theory.

In this paper, we first introduce the Bezoutian matrix and its application to
calculating the total sum of residues algebraically. Then we generalize this
method to rational integrands, with the help of an {\it elimination and
grevlex} monomial ordering. We apply this approach to the tree-level scattering
equations, to get the amplitudes without finding the explicit solutions or
summing over the residues. In particular, we emphasize that in all examples from
the scattering equations we have examined, the {\it dual form} from the
Bezoutian matrix computation has a strikingly simple form. This leads to a
shortcut which greatly enhances the amplitude calculation. Finally, we
explicitly show some high-multiplicity examples, like 8-point Yang-Mills
amplitudes and 10-point $\varphi^3$-theory amplitudes, to demonstrate the
strength of our method. 

{\bf Note:} During the preparation of this manuscript, an interesting preprint
by Feng, He, Huang and Rao \cite{Huang:2015yka} appeared. Clearly, the
motivation for these authors has been the same as ours, hence there is a natural
overlap with the present paper. We remark that the key difference between our
approach and ref.~\cite{Huang:2015yka} is that the principal object in our
paper---the Bezoutian matrix---is of size $(n-3)\times(n-3)$ only whereas the
companion matrix used in ref.~\cite{Huang:2015yka} grows factorially as
$(n-3)!\times(n-3)!$. Hence our method is able to efficiently evaluate
high-multiplicity (for example, 10-point) amplitudes. After this project was
completed, we were informed of another similar algorithm
\cite{Kalousios:2015fya}. The relation between refs.~\cite{Huang:2015yka} and
\cite{Kalousios:2015fya} has now been investigated \cite{Cardona:2015eba}.

\subsection{Examples of CHY Representations}
\label{SEC_CHY_EXAMPLES}
For later reference it is worthwhile to supply the reader with a few examples of
CHY representations of tree-level amplitudes. Pure Yang-Mills and
$\varphi^3$-theory give rise to CHY formulas that are quite representative to
the formalism and thus serve as appropriate testing grounds for our method. We
denote the canonically ordered $n$-gluon color-stripped amplitude as 
$\mc A_n^\tree$ and the $n$-point $\varphi^3$ amplitude as 
$\mc A_n^{\varphi^3,\tree}$.

The simplest instance of a CHY representation is found in massless
$\varphi^3$-theory. The prescription is simply to insert a squared Parke-Taylor
factor into the CHY integrand. In connection with the CHY construction, the
(canonically ordered) Parke-Taylor factor is understood as the following
expression,
\begin{align}
PT(1,2,\dots,n)\equiv\frac{1}{(z_1-z_2)(z_2-z_3)\cdots(z_n-z_1)}\;,
\end{align}
whose structure obviously bears resemblance with the Parke-Taylor denominator of
the maximally helicity violating (MHV) gluon tree-level amplitude. Amplitudes in
$\varphi^3$-theory can thus be written
\begin{align}
\mc A_n^{\varphi^3,\tree} = 
\oint_{\mc O}\frac{d\Omega_\CHY}{
(z_1-z_2)^2(z_2-z_3)^2\cdots(z_n-z_1)^2}\;.
\label{EQ_CHY_PHI3}
\end{align}
The representation of amplitudes in pure Yang-Mills is slightly more complicated
and involves in addition to a Parke-Taylor factor also a Pfaffian which encodes
the dependence of the external polarizations $\{\epsilon_i\}$. According to CHY
\cite{Cachazo:2013hca,Cachazo:2013iea},
\begin{align}
\mc A_n^\tree = {} &
\oint_{\mc O}d\Omega_\CHY
\frac{\Pf'\Psi(z,k,\epsilon)}{
(z_1-z_2)(z_2-z_3)\cdots(z_n-z_1)
}\;,
\label{EQ_CHY_YM}
\end{align}
where $\Psi$ is the $2n\times 2n$ antisymmetric matrix,
\begin{align}
\Psi = \left(
\begin{array}{cc}
A & -C^T \\
C & B
\end{array}
\right)\;.
\end{align}
The components of the $n\times n$ matrices $A$, $B$ and $C$ are as follows,
\begin{align}
A_{ab} = \left\{
\begin{array}{cl}
\dfrac{k_a\cdot k_b}{z_a-z_b} & \;\;a\neq b\;, \\[2mm]
0 & \;\;i = j\;,
\end{array}
\right.
\quad
B_{ab} = \left\{
\begin{array}{cl}
\dfrac{\epsilon_a\cdot\epsilon_b}{z_a-z_b} & \;\;a\neq b\;, \\[2mm]
0 & \;\;a = b\;,
\end{array}
\right.
\quad
C_{ab} = \left\{
\begin{array}{cl}
\dfrac{\epsilon_a\cdot k_b}{z_a-z_b} & \;\;a\neq b\;, \\[2mm]
-\sum_{c\neq a}\dfrac{\epsilon_a\cdot k_c}{z_a-z_c} & \;\;a = b\;.
\end{array}
\right.
\end{align}
Finally, the reduced Pfaffian in eq.~\eqref{EQ_CHY_YM} is defined in terms of
the matrix $\Psi_{ij}^{ij}$, obtained from $\Psi$ by removing rows $i,j$ and
columns $i,j$ for $1\leq i<j\leq n$, and is given by
\begin{align}
\Pf'\Psi\equiv 2\frac{(-1)^{i+j}}{(z_i-z_j)}\Pf\Psi_{ij}^{ij}\;.
\label{EQ_PFAFFIAN_PRIME}
\end{align}
Note that $\Pf'\Psi$ is independent of $i$ and $j$.

\subsection{The Polynomial Form of the Scattering Equations}
The scattering equations \eqref{EQ_CHY_EQS} proposed by Cachazo, He and Yuan are
rational; for our purposes, it is advantageous to reformulate them as polynomial
equations in order to enable systematic studies by means of computational
algebraic geometry. Here we follow Dolan and Goddard \cite{Dolan:2014ega}, who
recently proved that the $n$-point scattering equations, after properly fixing
the $\SL(2,\C)$ redundancy, are equivalent to a system of $(n-3)$ polynomial
equations,
\begin{align}
h_m = 0\;, \quad 1\leq m\leq n-3\;,
\end{align}
in $(n-3)$ variables. Each $h_m$ is a homogeneous polynomial of degree $m$ which
is linear in each variable taken separately. More precisely,
\begin{align}
h_m = \frac{1}{m!}\sum_{
\substack{
a_1,a_2,\dots,a_m\in A' \\
a_i\neq a_j}}
\sigma_{a_1a_2\cdots a_m}z_{a_1}z_{a_2}\cdots z_{a_m}\;,
\label{EQ_DOLAN_GODDARD}
\end{align}
where $A' = \{2,\dots,n-1\}$ and 
$\sigma_{a_1a_2\cdots a_m} = k_{1a_1a_2\cdots a_m}^2$. The polynomial equations
define a zero-dimensional projective algebraic variety in $\CP^{n-3}$ and the
number of solutions is $(n-3)!$ for generic kinematics by Bézout's theorem.

What remains is to reexpress the measure of the CHY formula \eqref{EQ_CHY_I} in
terms of the polynomials \eqref{EQ_DOLAN_GODDARD}. At this point it is
convenient to fix the $\SL(2,\C)$ redundancy and specialize to $z_1\to\infty$,
$z_2$ fixed and $z_n\to 0$, and also extract a Parke-Taylor factor from $\mc I$.
A short calculation shows that \cite{Dolan:2014ega}
\begin{align}
\mc A_n^\tree = \oint_{\mc O}d\tilde{\Omega}_\CHY\tilde{\mc I}(z,k)\;,
\label{EQ_CHY_DG}
\end{align}
where (up to an overall sign) 
\begin{align}
d\tilde{\Omega}_\CHY\equiv 
\frac{z_2}{z_{n-1}}
\prod_{m=1}^{n-3}\frac{1}{h_m(z,k)}
\prod_{2\leq a<b\leq n-1}\!(z_a-z_b)
\prod_{a=2}^{n-2}\frac{z_adz_{a+1}}{(z_a-z_{a+1})^2}\;.
\label{EQ_CHY_DG_MEASURE}
\end{align}
This formula completes our review of the scattering equations and CHY integrals.

\section{Multivariate Residues and the Bezoutian Matrix}
Motivated by the preceding discussion, we continue with a quick introduction to
the elementary theory of multivariate residues. For more details, refer to the
classical text books by Griffiths and Harris \cite{MR507725} and Hartshorne
\cite{MR0463157}.

\subsection{Local Residues}
Suppose that $f = (f_1,\dots,f_n) : \C^n\to\C^n$ is a holomorphic function with
an isolated common zero at $z = (z_1,\dots,z_n) = \xi\in\C^n$. Moreover, let
$N:\C^n\to\C$ be a meromorphic function and assume regularity of $N$ at 
$z = \xi$. Let $\omega$ be the $n$-form,
\begin{align}
\omega = \frac{N(z)}{f_1(z)\cdots f_n(z)}dz_1\wedge\cdots\wedge dz_n\;.
\label{EQ_DIFF_FORM}
\end{align}
The (local) residue of $\omega$ at $\xi$ with respect to the divisors
$\{f_1,\dots,f_n\}$ is by definition given by the multidimensional contour
integral,
\begin{align}
\Res_\xi(\omega) = \Res_{\{f_1,\dots,f_n\},\xi}(\omega) = 
\frac{1}{(2\pi i)^n}\oint_{\Gamma_\delta}
\frac{N(z)}{f_1(z)\cdots f_n(z)}dz_1\wedge\cdots\wedge dz_n\;,
\end{align}
where $\Gamma_\delta = \{z\in\C^n\,:\,|f_i(z)| = \delta_i\}$ is the real
$n$-dimensional cycle around $\xi$, oriented such that 
$d\arg f_1\wedge\cdots\wedge d\arg f_n\geq 0$. Here $\delta_i$ is a sufficiently
small positive number. Note that the residue is independent of the $\delta_i$s.

There are three types of multivariate residues: factorizable, nondegenerate and
degenerate. Factorizable residues are trivial to calculate: if 
$f_i(z) = f_i(z_i)$, the residue factorizes into a product of one-dimensional
contour integrals,
\begin{align}
\Res_{\{f_1,\dots,f_n\},\xi}(\omega) = 
\frac{1}{(2\pi i)^n}
\oint_{|f_1(z_1)|=\delta_1}\frac{dz_1}{f_1(z_1)}\cdots
\oint_{|f_n(z_n)|=\delta_n}\frac{dz_n}{f_n(z_n)}\;.
\end{align}
Generally speaking, the $f_i$'s are not univariate functions and to decide
whether the residue under consideration is degenerate or nondegenerate, we first
evaluate the Jacobian determinant,
\begin{align}
J(\xi)\equiv\det_{i,j}\bigg(
\pd{f_i}{z_j}\bigg)\bigg|_{z=\xi}\;.
\end{align}
If $J(\xi)$ is nonvanishing, the residue at $z = \xi$ is termed nondegenerate.
Under these circumstances the residue can easily be evaluated by applying
Cauchy's theorem in higher dimensions, with the result
\begin{align}
\Res_{\{f_1,\dots,f_n\},\xi}(\omega) = 
\frac{N(\xi)}{J(\xi)}\;.
\label{Cauchy}
\end{align}
Degenerate residues are more challenging and require algebraic geometry methods,
for example the transformation law~\cite{MR507725} or the Bezoutian
matrix~\cite{MR2161985}. Techniques for evaluating degenerate multivariate
residues are beyond the scope of the present paper. See instead applications in
the context of multiloop unitarity in
refs.~\cite{Sogaard:2013fpa,Sogaard:2014ila,Sogaard:2014oka,Johansson:2015ava}.

\subsection{Global Residues and the Bezoutian Matrix}
Often we are not interested in the details of each individual residue, but
rather the global structure of the total sum of the residues. For example, to
get tree amplitudes from the scattering equations, we only need to know the sum
of the $(n-3)!$ residues. It is frequently very difficult to calculate
individual residues, because complicated algebraic extensions appear in the
intermediate steps and the final result. Hence, we introduce an algebraic
geometry approach, the Bezoutian matrix method, which allows us to arrive at the
total sum of residues directly, without finding the singular point locus or
calculating individual residues. 

For the purposes of the remainder of this paper, it suffices to specialize to
situations involving only rational functions. More precisely, we consider
separately $n$ polynomials $\{f_1,\dots,f_n\}$ in the ring 
$R = \C[z_1,\dots,z_n]$ and an arbitrary numerator $N\in R$. We assume that 
the ideal\footnote{
The ideal $I$ generated by a set of polynomials $\{f_1,\dots,f_n\}$ is a special
subset of the ring $R = \C[z_1,\dots,z_n]$, defined as
$I = \avg{f_1,\dots,f_n}\equiv \{f\,|\, 
f = \sum_{i=1}^n a_if_i\,,\; a_i\in R\}$.
}
$I = \avg{f_1,\dots,f_n}$ is zero-dimensional, i.e. the zero locus $\mc Z(I)$
(set of all simultaneous zeros of the $f_i$s) consists only of a finite number
of points. Since $I$ is zero-dimensional, the quotient ring $R/I$ is a
finite-dimensional $\C$-linear space. 

In view of the above considerations, we simply define the {\it global residue}
as the sum of all the individual or local residues,
\begin{align}
\Res(N)\equiv\sum_{\xi_i\in\mc Z(I)}\Res_{\{f_1,\dots,f_n\},\xi_i}
\bigg(\frac{N(z)dz_1\wedge\cdots\wedge dz_n}{f_1(z)\cdots f_n(z)}\bigg)\;.
\label{EQ_GLOBAL_RES}
\end{align}
Stokes' theorem ensures that the values of the residues only depend on the
equivalence class $[N]$ of $N$ in $R/I$~\cite{MR507725}. This equivalence splits
polynomials into subclasses of polynomials which have the same remainder after
polynomial division. Unlike the familiar division algorithm for univariate
polynomials, multivariate polynomial division is only well-defined if performed
towards a Gr{\"o}bner basis (in some monomial order) of the ideal $I$. (For the
benefit of the non-expert reader, we remark that a Gr{\"o}bner basis is a
particular set of generators of an ideal. The theory of Gr{\"o}bner bases is
perhaps the most important practical tool in computational algebraic geometry;
it allows for e.g. non-linear generalization of Gaussian elimination and
multivariate polynomial division.) Then for any $N\in R$, there is a unique
polynomial $r\in R$ (the unique remainder, for the monomial order being
considered), which yields a representation of the equivalence class 
$[N]\in R/I$) such that $N(z) = q(z)+r(z)$ with $q\in I$. In particular, if $N$
coincidentally belongs to $I$, the residue vanishes identically. 

The mathematical question we would now like to pose is how to obtain the value
of the global residue \eqref{EQ_GLOBAL_RES}, without completing the instructed
summation over the local residues. Given a pair of polynomials $N_1,N_2$ we can
introduce a symmetric inner product,
\begin{align}
\avg{N_1,N_2}\equiv\Res(N_1\cdot N_2)\;.
\end{align}
The following theorem serves as the heart of computations of this paper.
\begin{thm}[Global Duality]
$\avg{\bullet,\bullet}$ is a nondegenerate inner product in $R/I$.
\end{thm}
The proof is omitted here, but can be found in ref.~\cite{MR507725}. Now suppose
that $\{e_i\}$ forms a linear basis of $R/I$. This merely means that any
remainder can be written relative to the monomials $\{e_i\}$. This theorem
implies the existence of a {\it dual basis} $\{\Delta_i\}$ in $R/I$, with the
property,
\begin{align}
\avg{e_i,\Delta_j} = \delta_{ij}\;.
\end{align}
The virtue of the dual basis is that it characterizes the structure of global
residues in an explicit manner, without reference to the individual residues or
their locations. To realize this, expand the remainder $[N]$ over the canonical
linear basis,
\begin{align}
[N] =\sum_i\lambda_i e_i\;,\quad \lambda_i\in\C\;,
\end{align}
and similarly, decompose unity using the dual basis,
\begin{align}
\label{partition}
1 = \sum_i\mu_i\Delta_i\;, \quad \mu_i\in\C\;.
\end{align}
Then, by construction, the global residue is given by \cite{MR2161985}
\begin{align}
\label{sum}
\Res(N) =\avg{N,1} = 
\sum_{i,j}\lambda_i\mu_j\avg{e_i,\Delta_j} = \sum_i\lambda_i\mu_i\;.
\end{align}
In particular, if one term in the dual basis, $\Delta_s$, is a constant, then
the global residue computation is extremely simple,
\begin{equation}
  \label{short_cut}
  \Res(N) =\avg{N,\Delta_s}/\Delta_s=\lambda_s/\Delta_s\;.
\end{equation}
The partition \eqref{partition} and summation \eqref{sum} are not needed for
this case. We have explicitly verified that the tree-level scattering equations
have this remarkable feature for at least up to $n = 10$.

The canonical linear basis and the dual basis can be obtained in practice by
means of the Gr{\"o}bner basis method and the Bezoutian matrix \cite{MR2161985}.
The procedure is described in the following.
\begin{enumerate}
\item (Canonical Linear Basis of Quotient Ring)
Calculate the Gr{\"o}bner basis of $I$, $G$, in Degree Lexicographic (grlex) or
Degree Reverse Lexicographic (grevlex) order. Identify the leading terms $LT(G)$
for all polynomials in $G$. The canonical linear basis, $\{e_i\}$, for $R/I$
consists of all monomials in $R$ which are lower than $LT(G)$, with respect to
the monomial order.
\item (Dual Basis of Quotient Ring)
Define the $n\times n$ Bezoutian matrix $B$ for $I$,
\begin{align}
B_{ij}(z,y)\equiv\frac{
f_i(y_1,\dots,y_{j-1},z_j,\dots,z_n)-
f_i(y_1,\dots,y_j,z_{j+1},\dots,z_n)}{z_j-y_j}\;.
\end{align}
Calculate the determinant of $B$, $\det B$. Define $\tilde G$ as $G$ subject to
the replacement $z_i\to y_i$ for $i = 1,\dots,n$. Carry out the multivariate
polynomial division of $\det B$ over $G\otimes\tilde G$ and obtain the
remainder,
\begin{align}
\sum_i a_i(y)e_i(z)\;.
\end{align}
The dual basis $\{\Delta_i\}$ for $\{e_i\}$, with respect to
$\avg{\bullet,\bullet}$, is $\Delta_i = a_i(z)$ \cite{MR2161985}.
\end{enumerate}

At this moment it is important to underline that all operations related to
Gr{\"o}bner bases require a certain choice of monomial ordering, for example
Lexicographic (lex), Degree Lexicographic (grlex), Degree Reverse Lexicographic
(grevlex). Moreover, the choice of monomial ordering may drastically influence
on the speed of the calculation. Although some of these monomial orderings are
rather self-explanatory, we supply here a concise clarification. Consider the
shorthand notation for monomials, 
$z^\alpha\equiv z_1^{\alpha_1}\cdots z_n^{\alpha_n}$.
\begin{itemize}
\item $z^\alpha\succ_{\text{lex}}z^\beta$ if the left-most nonzero entry of
$\alpha-\beta$ is positive. It follows that there are $n!$ nonequivalent
lexicographic orderings, corresponding to the particular orderings of the
variables. This monomial ordering is often slow to use in connection with
Gr{\"o}bner bases.
\item $z^\alpha\succ_{\text{grlex}}z^\beta$ if $|\alpha| > |\beta|$, or if 
$|\alpha| = |\beta|$ and $z^\alpha\succ_{\text{lex}}z^\beta$. In words, the
grlex order first compares total degree and then applies the lexicographic
order.
\item $z^\alpha\succ_{\text{grevlex}}z^\beta$ if $|\alpha| > |\beta|$, or if 
$|\alpha| = |\beta|$ and the right-most nonzero entry of $\alpha-\beta$ is
negative. We stress that this monomial order is often the most efficient order
for constructing Gr{\"o}bner bases.
\end{itemize}

We conclude this subsection with a quick and transparent example of how the
Bezoutian matrix can be used in practice to calculate global residues, without
the need for obtaining the individual residues.

\begin{example}[Global Residues and the Bezoutian Matrix]
\normalfont
\label{EX_BEZOUTIAN}
Let $R = \C[z_1,z_2]$ and consider the zero-dimensional ideal 
$I = \avg{2-z_1-z_2,z_1+z_2+2z_1z_2}\subset R$. The zero locus is clearly 
$\mc Z(I) = \{(1-\sqrt{2},1+\sqrt{2}),(1+\sqrt{2},1-\sqrt{2})\}$. The
Gr{\"o}bner basis of $I$ in grlex order is 
$G = \{z_1+z_2-2,1+2z_2-z_2^2\}$ and therefore the canonical linear basis of
$R/I$ is $\{e_i\} = \{z_2,1\}$. The Bezoutian matrix takes the form
\begin{align}
B = \left(
\begin{array}{cc}
-1 & -1 \\
1+2z_2 & 1+2y_1
\end{array}
\right)\;,
\end{align} whence $\det B = 2(z_2-y_1)$. Polynomial division yields the dual
basis $\{\Delta_i\} = \{2,2(z_2-2)\}$ and since $1 = \frac{1}{2}\Delta_1$ we
have $\{\mu_i\} = \big\{\frac{1}{2},0\big\}$. Let us now pick a numerator, say,
$N(z) = z_2^2$; for this choice the decomposition over the canonical linear
basis is $\{\lambda_i\} = \{2,1\}$. The global residue thus takes the value
\begin{align}
\Res(N) = \sum_{i=1,2}\lambda_i\mu_i = 1
\end{align}
and the result matches the sum of the two individual residues,
\begin{align}
\Res(N) =  
\frac{1}{8}\big(4+3\sqrt{2}\big)+\frac{1}{8}\big(4-3\sqrt{2}\big) = 1\;.
\end{align}
Note that by the Bezoutian matrix computation, we do not need the algebraic
extension from $\sqrt{2}$. This greatly simplifies computation for more
complicated examples. 
\end{example}

\subsection{Our Proposal for CHY Integrals}
The Bezoutian matrix method provides us with a highly efficient technique for
computing global residues of differential forms of the kind
eq.~\eqref{EQ_DIFF_FORM}, where the numerator and denominator factors are
polynomials, without the need for calculating the local residues individually.
The only obstacle to immediately apply the theory of global residues in
connection with the polynomial scattering equations \eqref{EQ_DOLAN_GODDARD} and
the CHY representation \eqref{EQ_CHY_DG} is the presence of extra denominator
factors, for example the Parke-Taylor factors. Phrased slightly differently, the
numerator $N$ in our problem is not a polynomial, but a rational function.

The trick is to replace the extra denominator by its polynomial inverse in $R/I$
in the numerator. Let $N=h/g$ where $h,g\in R$. For a finite residue, $g$ should
not vanish on $\mc Z(I)$ so $\{f_1,\ldots f_n,g\}$ have no common zero. By
Hilbert's Nullstellensatz, there exist polynomials 
$a_1,\dots,a_n,\tilde g \in R$ such that,
\begin{equation}
  a_1 f_1+\cdots+a_n f_n+\tilde g g=1\;.
\label{Hilbert_Nullstellensatz}
\end{equation}
Suppose that all residues of $I$ are nondegenerate. Then by
\eqref{Hilbert_Nullstellensatz},
\begin{eqnarray}
  \label{eq:2}
\Res_{\xi}(N) =\Res_{\xi}(h \tilde g) + \sum_{i=1}^n
  \Res_{\xi}(a_if_ih/g)=\Res_{\xi}(h \tilde g)\;, 
\end{eqnarray}
because $a_if_ih/g$ is in the ideal generated by the $f_i$'s in the ring of germs
of holomorphic functions around $\xi$~\cite{MR507725}. So for residue
computations, we are free to replace $g^{-1}$ by the polynomial $\tilde g$. If
$g$ factorizes, the polynomial inverse of $g$ equals the product of the
inverses. This elementary observation typically greatly simplifies the problem,
especially for the scattering equations. 

Consequently, the rational numerator is converted to a purely polynomial form.
The above discussion is implemented in Mathematica using Macaulay2 \cite{M2} via
the MathematicaM2\footnote{The converting matrix of the Gr{\"o}bner basis is not
directly provided by Mathematica.} package \cite{MathematicaM2}.

\begin{algorithm}[Polynomial Inverse]
\normalfont
\label{ALG_POLY_INVERSE}
Let $I = \avg{f_1,\dots,f_n}\subset \C[z]$ be a zero-dimensional ideal and 
suppose\footnote{That is to say, the polynomials $\{f_1,\dots,f_n,g\}$ have no
simultaneous zero.} $\avg{g}+I=R$. Calculate the generator of the Gr{\"o}bner
basis of the ideal $\avg{f_1,\dots,f_n,g}$ in some monomial order and record the
converting matrix, so that $1 = a_1 f_1+\cdots+a_n f_n+\tilde g g$. The
polynomial inverse of $g$, with respect to $I$, is $\tilde g$.
\end{algorithm}

However, this algorithm requires the converting matrix for the Gr{\"o}bner basis
computation. In some cases, it is time and memory consuming. Therefore we may
use a more efficient algorithm.
\begin{algorithm}[Polynomial Inverse, Enhanced]
\normalfont
\label{ALG_POLY_INVERSE2} Let $I = \avg{f_1,\dots,f_n}\subset \C[z]$ be a
zero-dimensional ideal and suppose $\avg{g}+I=R$. Introduce the auxiliary
variable $w$ and the ideal $J=\avg{f_1,\dots,f_n,w g-1}$ in 
$\C[w,z_1,\dots,z_n]$. Define a monomial order $\mc T$ that (1) compares the
degrees of $w$ first and (2) applies grevlex order for $\{z_1,\dots,z_n\}$
as $z_1\succ\cdots\succ z_n$. Then calculate the Gr{\"o}bner basis $G(J)$ of $J$
in the order of $\mc T$. Inside $G(J)$, there must be a polynomial linear in
$w$,
\begin{equation}
  \label{eq:1}
  w - \tilde g(z_1,\dots,z_n) \in G(J)\;.
\end{equation}
Then $\tilde g(z_1,\dots,z_n)$ is the inverse of $g$, with respect to $I$.
\end{algorithm}
The correctness of the output is guaranteed by the definition of $\mc T$. This
algorithm does not need the converting matrix. Fast routines for Gr{\"o}bner
basis computations, like Faug\`ere's $F4$ and $F5$ algorithms \cite{F4,F5}, can
be applied. In practice, we use the FGb package \cite{FGb}.

\begin{example}[Polynomial Inverses and Global Residues]
\normalfont
For brevity we will merely revisit the problem in Example~\ref{EX_BEZOUTIAN},
but now with a rational function $N(z) = 1/z_1^2$ in the numerator. The
corresponding polynomial inverse in $R/I$ is quickly calculated using Algorithm
\ref{ALG_POLY_INVERSE}. The Gr{\"o}bner basis computation gives
\begin{equation}
  \label{eq:3}
  1=\frac{1+2z_1+2z_2+4z_1z_2}{2} (2-z_1-z_2)
  +\frac{-3+2z_2}{2}(z_1+z_2+2 z_1 z_2)+(1+2z_2)(z_1^2)\;.
\end{equation}
It is immediately clear that the inverse of $1/z_1^2$ with respect to $I$ is
$1+2z_2$. Following the steps of Example~\ref{EX_BEZOUTIAN} determines the
global residue to be $1$.

Alternatively we can apply Algorithm~\ref{ALG_POLY_INVERSE2}. Introduce the
auxiliary variable $w$, and define $J=I+\avg{w z_1^2-1}$. Considering the
ordering $\mc T$, the Gr{\"o}bner basis is
\begin{equation}
  \label{eq:5}
  G(J)=\{-2+z_1+z_2,1-2z_2+z_2^2,w-1-2z_2\}\;.
\end{equation}
Hence the inverse of $1/z_1^2$ with respect to $I$ is $1+2z_2$.
\end{example}

\section{Examples}
We present several explicit examples of how to employ the Bezoutian matrix
method to evaluate scattering amplitudes in the CHY formalism. Without loss of
the main features we will primarily be interested in massless $\varphi^3$-theory
and pure Yang-Mills. In the beginning we consider analytic calculations for
lower multiplicity kinematics. As the complexity of the intermediate and final
analytic expressions increases significantly with the number of particles, it is
instructive to analyze higher-point examples using numerical data for the
kinematic invariants.

\subsection{Four-Point Amplitudes}
There is only one independent scattering equation for four external particles
and thus a single univariate residue to compute. The pole in $z_3$ is trivial to
locate,
\begin{align}
h_1 = \sigma_2z_2+\sigma_3z_3 = 0 \Lra
\frac{z_3}{z_2} = -\frac{\sigma_2}{\sigma_3} = -\frac{s_{12}}{s_{13}}\;.
\end{align}
We can now effortlessly extract the four-scalar amplitude from the CHY
representation \eqref{EQ_CHY_DG}, with the familiar result 
\begin{align}
\mc A_4^{\varphi^3,\tree} = 
-\oint_{\mc O}
\frac{dz_3}{\sigma_2z_2+\sigma_3z_3}\frac{z_2^2}{(z_2-z_3)z_3} =
\frac{1}{s_{12}}+\frac{1}{s_{14}}\;.
\end{align}

Even though this example is very simple, let us for the sake of
completeness work it out using the Bezoutian matrix approach. We readily arrive
at $\{e_i\} = \{1\}$, $\det B = \sigma_3$ and $\{\Delta_i\} = \{\sigma_3\}$.
Denote $g_1 = z_2-z_3$ and $g_2 = z_3$. Then by Algorithm~\ref{ALG_POLY_INVERSE},
\begin{align}
\tilde g_1 = \frac{\sigma_3}{(\sigma_2+\sigma_3)z_2}\;, \quad
\tilde g_2 = -\frac{\sigma_3}{\sigma_2z_2}\;,
\end{align}
and therefore, as expected,
\begin{align}
\mc A_4^{\varphi^3,\tree} = 
\oint_{\mc O}
\frac{dz_3}{\sigma_2z_2+\sigma_3z_3}
\frac{\sigma_3^2}{\sigma_2(\sigma_2+\sigma_3)} =
\frac{\sigma_3}{\sigma_2(\sigma_2+\sigma_3)}\;,
\end{align}
where in the last step we invoked eq.~\eqref{short_cut}.

\subsection{Five-Point Amplitudes}
The five-particle case gives rise to the first nontrivial instance of the
scattering equations and provides a prime example of the intermediate steps of
the proposed method. We examine the two independent polynomial scattering
equations,
\begin{align}
h_1 = {} & \sigma_2z_2+\sigma_3z_3+\sigma_4z_4 = 0\;, \nn \\ 
h_2 = {} & \sigma_{23}z_2z_3+\sigma_{24}z_2z_4+\sigma_{34}z_3z_4 = 0\;, 
\end{align}
whose two solutions we quote for later reference (setting $z_2 = 1$ for
simplicity),
\begin{align}
\mc S_1:\, {} &
\left\{
\rule{0cm}{1cm}
\right.
\begin{array}{l}
z_3 =
-\dfrac{\sigma_2\sigma_{34}+\sigma_3\sigma_{24}
-\sigma_4\sigma_{23}+\sqrt{\Delta}}{2\sigma_3\sigma_{34}}\;, \\[1mm]
z_4 =
-\dfrac{\sigma_2\sigma_{34}-\sigma_3\sigma_{24}
+\sigma_4\sigma_{23}-\sqrt{\Delta}}{2\sigma_3\sigma_{34}}\;,
\end{array}
\label{EQ_H1H2_S1}
\\[1mm]
\mc S_2:\, {} &
\left\{
\rule{0cm}{1cm}
\right.
\begin{array}{l}
z_3 =
-\dfrac{\sigma_2\sigma_{34}+\sigma_3\sigma_{24}
-\sigma_4\sigma_{23}-\sqrt{\Delta}}{2\sigma_3\sigma_{34}}\;, \\[1mm]
z_4 =
-\dfrac{\sigma_2\sigma_{34}-\sigma_3\sigma_{24}
+\sigma_4\sigma_{23}+\sqrt{\Delta}}{2\sigma_3\sigma_{34}}\;.
\end{array}
\label{EQ_H1H2_S2}
\end{align}
The discriminant is given by
\begin{align}
\Delta =
(\sigma_2\sigma_{34}+\sigma_3\sigma_{24}-\sigma_4\sigma_{23})^2
-4\sigma_2\sigma_3\sigma_{24}\sigma_{34}\;.
\end{align}
According to the CHY prescription \eqref{EQ_CHY_DG}, the five-point amplitude
in $\varphi^3$-theory is computed by the two-dimensional contour integral,
\begin{align}
\mc A_5^{\varphi^3,\tree} = \oint_{\mc O}
\frac{dz_3dz_4}{h_1h_2}\frac{z_3(1-z_4)}{(1-z_3)(z_3-z_4)z_4}\;,
\end{align}
where the contour $\mc O$ encloses the two simultaneous zeros
\eqref{EQ_H1H2_S1}-\eqref{EQ_H1H2_S2} of $h_1$ and
$h_2$, but no other singularities. The denominator of the integrand contains
three additional linear factors $g_1 = 1-z_3$, $g_2 = z_3-z_4$ and $g_3 = z_4$.
Let $I = \avg{h_1,h_2}\subset R = \C[z_3,z_4]$. Using
Algorithm~\ref{ALG_POLY_INVERSE} it takes Mathematica only a split second to
find the following expressions for the polynomial inverses,
\begin{align}
\tilde g_1 = {} &
\frac{\sigma_3\sigma_{24}+\sigma_3\sigma_{34}
-\sigma_4(\sigma_{23}+\sigma_{34}z_4)}{
(\sigma_2+\sigma_3)(\sigma_{24}+\sigma_{34})-\sigma_4\sigma_{23}}\;, \\[1mm]
\tilde g_2 = {} &
-\frac{\sigma_2\sigma_3\sigma_{34}-(\sigma_3+\sigma_4)(
\sigma_3\sigma_{24}-\sigma_4(\sigma_{23}+\sigma_{34}z_4))}{
\sigma_2(\sigma_2\sigma_{34}-(\sigma_3+\sigma_4)(\sigma_{23}+\sigma_{24}))}\;,
\\[1mm]
\tilde g_3 = {} &
\frac{\sigma_3(\sigma_{24}+\sigma_{34}z_3)
-\sigma_4\sigma_{23}}{\sigma_2\sigma_{23}}\;.
\end{align}
The amplitude can then be rewritten in terms of a polynomial numerator function
$N$,
\begin{align}
\mc A_5^{\varphi^3,\tree} =
\oint_{\mc O}dz_3dz_4\frac{N}{h_1h_2}\;, \quad
N(z_3,z_4) = z_3(1-z_4)\prod_{i=1}^3\tilde g_i\;.
\end{align}
The Gr{\"o}bner basis of $I$ in the grlex order $z_3\succ z_4$ is
\begin{align}
G = \{\sigma_2+\sigma_3z_3+\sigma_4z_4,\sigma_3\sigma_{24}z_4-
(\sigma_{23}+\sigma_{34}z_4)(\sigma_2+\sigma_4z_4)\}
\end{align}
whence the quotient ring basis $\{p_i\} = \{z_4,1\}$ of $R/I$ can be read off
immediately. Now, from the Bezoutian matrix,
\begin{align}
B(z,y) = \left(
\begin{array}{cc}
\sigma_3 & \sigma_4 \\
\sigma_{23}+\sigma_{34}z_4 & \sigma_{24}+\sigma_{34}y_3
\end{array}
\right)\;,
\end{align}
we calculate the determinant,
\begin{align}
\det B = \sigma_3(\sigma_{24}+\sigma_{34}y_3)
-\sigma_4(\sigma_{23}+\sigma_{34}z_4)\;,
\end{align}
and thus by polynomial division over $G\otimes\tilde G$, we derive the dual basis,
\begin{align}
\{\Delta_i\} = 
\{-\sigma_4\sigma_{34}, -\sigma_2\sigma_{34}+\sigma_3\sigma_{24}
-\sigma_4(\sigma_{23}+\sigma_{34}z_4)\}\;. 
\end{align}
The fact that the dual basis $\{\Delta_i\}$ contains a constant term immediately
allows us to exploit eq.~\eqref{short_cut} once we have obtained the
$z_4$-coefficient of $[N]$. The global residue and therefore the amplitude in
question reduce to a single term. Expressed as a function of the
$\sigma$-invariants, we arrive at the final result,
\begin{align}
\mc A_5^{\varphi^3,\tree} =
\frac{
\sigma_2\sigma_4\sigma_{23}\sigma_{24}
-\big(\sigma_3+\sigma_4\big)\big(
\sigma_3\left(\sigma_{24}+\sigma_{34}\right)-\sigma_4\sigma_{23}\big)
\sigma_{24}
+\sigma_2\sigma_3\big(\sigma_{23}+\sigma_{34}\big)
\big(\sigma_{24}+\sigma_{34}\big)}{
\sigma_2\sigma_{23}\big(
\sigma_2\sigma_{34}-\big(\sigma_3+\sigma_4\big)\big(
\sigma_{23}+\sigma_{24}\big)\big)\big(\big(\sigma_2+\sigma_3\big)
\big(\sigma_{24}+\sigma_{34}\big)-\sigma_4\sigma_{23}\big)}
\;.
\end{align}

The physical singularities of the five-point amplitude are the five independent
Mandelstam invariants $s_{12}$, $s_{23}$, $s_{34}$, $s_{45}$ and $s_{51}$.
Rewriting the $\sigma$-variables using simple kinematic identities immediately
leads to the well known Feynman diagram result, 
\begin{align}
\mc A_5^{\varphi^3,\tree} = 
\frac{1}{s_{12}s_{34}}+
\frac{1}{s_{12}s_{45}}+
\frac{1}{s_{23}s_{51}}+
\frac{1}{s_{23}s_{45}}+
\frac{1}{s_{34}s_{51}}\;.
\label{EQ_PHI3_5PT}
\end{align}
Direct evaluation of the individual residues of course yields the same answer.
However, the symbolic manipulations and cancellations are quite involved due to
the presence of square roots in eqs.~\eqref{EQ_H1H2_S1} and \eqref{EQ_H1H2_S2}.
Indeed, for comparison, we have the intermediate result
\begin{align}
\mc A_5^{\varphi^3,\tree} = \oint_{\mc O}
\frac{dz_3dz_4}{h_1h_2}\frac{z_3(1-z_4)}{(1-z_3)(z_3-z_4)z_4} = 
\Res_1+\Res_2\;,
\end{align}
where the residues are the rather complicated expressions,
\begin{align}
\Res_1 = {} &
-\frac{
2\sigma_3\sigma_4\sigma_{34}\big(
\sigma_2\sigma_{34}+\sigma_3\sigma_{24}-\sigma_4\sigma_{23}+\sqrt{\Delta}
\big)}{
\sqrt{\Delta}\big[\big(\sigma_3+\sigma_4\big)
\big(
\sigma_3\sigma_{24}-\sigma_4\sigma_{23}+\sqrt{\Delta}\big)
-\sigma_2\big(\sigma_3-\sigma_4\big)\sigma_{34}\big]
}
\\ & \qquad\;\times
\frac{
\sigma_3\sigma_{24}-\sigma_4\sigma_{23}
-\big(\sigma_2+2\sigma_4\big)\sigma_{34}+\sqrt{\Delta}
}{
\big(
\sigma_2\sigma_{34}-\sigma_3\sigma_{24}+\sigma_4\sigma_{23}-\sqrt{\Delta}
\big)
\big(
\sigma_3\sigma_{24}-\sigma_4\sigma_{23}+\big(\sigma_2+2\sigma_3
\big)\sigma_{34}+\sqrt{\Delta}
\big)
}\;,
\nn \\[1mm]
\Res_2 = {} &
+\frac{
2\sigma_3\sigma_4\sigma_{34}\big(
\sigma_2\sigma_{34}+\sigma_3\sigma_{24}-\sigma_4\sigma_{23}-\sqrt{\Delta}
\big)}{
\sqrt{\Delta}\big[\big(\sigma_3+\sigma_4\big)
\big(
\sigma_3\sigma_{24}-\sigma_4\sigma_{23}-\sqrt{\Delta}\big)
-\sigma_2\big(\sigma_3-\sigma_4\big)\sigma_{34}\big]
}
\\ & \qquad\;\times
\frac{
\sigma_3\sigma_{24}-\sigma_4\sigma_{23}
-\big(\sigma_2+2\sigma_4\big)\sigma_{34}-\sqrt{\Delta}
}{
\big(
\sigma_2\sigma_{34}-\sigma_3\sigma_{24}+\sigma_4\sigma_{23}+\sqrt{\Delta}
\big)
\big(
\sigma_3\sigma_{24}-\sigma_4\sigma_{23}+\big(\sigma_2+2\sigma_3
\big)\sigma_{34}-\sqrt{\Delta}
\big)
}\;. \nn
\end{align}

\subsection{Eight-Point Amplitudes}
We will now illustrate and validate the Bezoutian matrix method in a more
difficult situation, namely for an eight-gluon amplitude in pure Yang-Mills
theory. For the sake of simplicity we will restrict to four dimensions and
numerically study the MHV configuration where two of the gluons $i$ and $j$ have
negative helicity and the rest have positive helicity. The result is thus
straightforward to compare with the known answer due to the Parke-Taylor
formula,
\begin{align}
\mc A^\tree_{n,ij} = 
\frac{\spaa{i}{j}^4}{\spaa{1}{2}\spaa{2}{3}\cdots\spaa{n}{1}}\;.
\label{EQ_PARKE_TAYLOR}
\end{align}
High-multiplicity kinematics is conveniently generated through momentum
twistors. For this example we will proceed with the following values for the
momenta,
\begin{align}
k_1^\mu = {} &
\big(5/2,5/2,-i/2,1/2\big)\;, &
k_2^\mu = {} &
\big(-3/4,3/4,-3i/4,-3/4\big),\; \nn \\
k_3^\mu = {} &
\big(-1/6,0,0,1/6\big)\;, &
k_4^\mu = {} &
\big(-19/84,11/28,9i/28,1/84\big)\;, \nn \\
k_5^\mu = {} &
\big(10/21,-55/42,95i/42,40/21\big)\;, &
k_6^\mu = {} &
\big(23/12,-1/12,-i/12,23/12\big)\;, \nn \\
k_7^\mu = {} & 
\big(-19/28,1/28,-43i/28,-47/28\big)\;, &
k_8^\mu = {} &
\big(-43/14,-16/7,2i/7,-29/14\big)\;,
\end{align}
for which the amplitude \eqref{EQ_PARKE_TAYLOR} becomes, 
\begin{align}
\mc A_8^\tree(1^-,2^-,3^+,\dots,8^+) = \frac{4}{441}\;.
\label{EQ_PARKE_TAYLOR_VALUE}
\end{align}

At eight points, there are five independent scattering equations $h_m = 0$,
$1\leq m\leq 5$, in the variables $z_3,\dots,z_7$, with $(8-3)! = 120$
simultaneous solutions. Here we have gauge fixed $z_1\to\infty$, $z_2\to 1$ and
$z_8\to0$ as usual. Needless to say, it is almost impossible to compute the
desired amplitude in practice by evaluating the sum over the individual
residues. The scattering equations are a bit lengthy for problems with many
particles, but otherwise elementary to write down explicitly. 

As previously explained in Section~\ref{SEC_CHY_EXAMPLES}, Yang-Mills amplitudes
in the CHY representation involve a Pfaffian. Recall that for the CHY formula
with polynomial denominators, we pulled out a Parke-Taylor factor from the CHY
integrand in eq.~\eqref{EQ_CHY_DG}. Therefore, the gauge fixed Yang-Mills
integrand is given by the limit,
\begin{align}
\tilde{\mc I}(z,k) = 
\prod_{a=2}^{n-1}(z_a-z_{a+1})
\lim_{z_1\to\infty}(z_1^2\Pf'\Psi)\;.
\end{align}
The numerator of $\tilde{\mc I}$ is a huge polynomial whose explicit form is not
particularly important for exposing the essential steps of the calculation. On
the other hand, the denominator of $\tilde{\mc I}$ is the simple polynomial
function (up to an overall constant),
\begin{gather}
D_{YM} = z_3z_4(z_3-z_5)(1-z_4)z_5(1-z_5)(z_3-z_6) \nn \\
(z_4-z_6)(1-z_6)z_6(z_3-z_7)(z_4-z_7)(z_5-z_7)(1-z_7)\;.
\end{gather}
Moreover, there is a contribution to the denominator coming from
$d\tilde{\Omega}_\CHY$ \eqref{EQ_CHY_DG_MEASURE},
\begin{align}
D_{DG} = (z_2-z_3)(z_3-z_4)(z_4-z_5)(z_5-z_6)(z_6-z_7)z_7\;.
\end{align}

We invert the polynomial factors of these denominators via
Algorithm~\ref{ALG_POLY_INVERSE2}. The Gr{\"o}bner basis computations are
rapidly performed using the FGb library \cite{FGb} in Maple. For the dual basis
$\{\Delta_i\}$, determined by applying the Bezoutian matrix method, we
explicitly observe that $\Delta_1$ is a constant, so we can use
eq.~\eqref{short_cut}. The corresponding term in the canonical linear basis 
$e_1 = z_7^{10}$. Consequently,
\begin{align}
\mc A_8^\tree(1^-,2^-,3^+,\dots,8^+) = 
\bigg[\frac{N_{DG}}{D_{DG}}\frac{N_{YM}}{D_{YM}}\bigg]_{z_7^{10}}/\Delta_1\;,
\end{align}
where the subscript indicates that we only take the coefficient for $z_7^{10}$.
The square brackets refer to the canonical form with respect to $I$, in grevlex
order. The end result of this calculation agrees with the value in
eq.~\eqref{EQ_PARKE_TAYLOR_VALUE}.

\subsection{Ten-point Amplitudes}
To fully demonstrate the power of our method, we calculate as a final example
the $10$-point $\varphi^3$ amplitude in four dimensions by applying the
Bezoutian matrix and the enhanced inversion algorithm
(Algorithm~\ref{ALG_POLY_INVERSE2}). Using momentum twistors, we consider the
numeric phase point,
\begin{align}
k_1^\mu = {} &
\big(6/5,6/5,-3i/5,-3/5\big)\;, &
k_2^\mu = {} &
\big(7/4,3/4,9i/4,-11/4\big),\; \nn \\
k_3^\mu = {} &
\big(-5/4,-5/4,-15i/4,15/4\big)\;, &
k_4^\mu = {} &
\big(-1,1/8,i/8,-1\big)\;, \nn \\
k_5^\mu = {} &
\big(1/10,-5/8,47i/40,-1\big)\;, &
k_6^\mu = {} &
\big(17/5,-7/2,-13i/10,1\big)\;, \\
k_7^\mu = {} & 
\big(-2,3,3i,2\big)\;, &
k_8^\mu = {} &
\big(-1,5,-7i,-5\big)\;, \nn \\
k_9^\mu = {} &
\big(-11/4,-6,6i,11/4\big)\;, &
k_{10}^\mu = {} &
\big(31/20,13/10,i/10,17/20\big)\;. \nn
\end{align}
There are $7$ scattering equations in the variables $z_1,\dots,z_7$. The
quotient ring $R/I$ has the dimension $(10-3)!= 5040$. In this case, Gr{\"o}bner
basis computation is heavy. So we use the fast Gr{\"o}bner basis computation
package \cite{FGb}. Furthermore, to reduce memory usage, we apply the finite
field technique: 
\begin{enumerate}
\item Calculate the global residue with the coefficients in the finite
  field $\Z/p$, where $p$ is a prime number. Repeat this process
  several times for different prime numbers. For this particular
  amplitude, we find that it is enough to use $12$ prime numbers, each of
  which is around the
  order of $10^4$.
\item Use the modular method \cite{Farey} to lift the global residue evaluated in
  finite fields, to a rational number, i.e. the physical value. 
\end{enumerate}

We calculate the dual basis, $\{\Delta_i\}$, in grevlex order, for $I$ via
the Bezoutian matrix method. For this example we also find that one of the dual
basis terms $\Delta_1$ is constant. The corresponding term in canonical basis is
$e_1=z_9^{21}$. By eq.~\eqref{short_cut},
\begin{equation}
  \label{eq:6}
  \mc A_{10}^{\varphi^3,\tree}=\bigg[\frac{N}{D}\bigg]_{z_9^{21}}/\Delta_1\;,
\end{equation}
where
\begin{gather}
  \label{eq:7}
  N=z_3 (1 - z_4) z_4 (1 - z_5) (z_3 - z_5) z_5 (1 - z_6) (z_3 - z_6) (z_4 - 
   z_6) z_6 (1 - z_7)  \nonumber\\ (z_3 - z_7) (z_4 - z_7) (z_5 - z_7) z_7 (1 - z_8) (z_3 - 
   z_8) (z_4 - z_8) (z_5 - z_8) (z_6 - z_8)  \nonumber\\ z_8 (1 - z_9) (z_3 - z_9) (z_4 - 
   z_9) (z_5 - z_9) (z_6 - z_9) (z_7 - z_9)\;,
\end{gather}
and
\begin{gather}
  D=(1 - z_3) (z_3 - z_4) (z_4 - z_5) (z_5 - z_6) (z_6 - z_7) (z_7 - z_8) (z_8 -
z_9) z_9\;.
\end{gather}
Again, we invert the factors in $D$ one by one to get a purely polynomial form
of the remaining integrand. The final result for this phase point is
\begin{equation}
  \label{eq:8}
  \mc A_{10}^{\varphi^3,\tree}
  =-\frac{248907703337666902407787}{24536182021587817097932800}\;.
\end{equation}

\section{Conclusion}
In summary we have employed computational algebraic geometry to study the
polynomial form of the scattering equations and to calculate various amplitudes
from CHY representations of pure Yang-Mills and $\varphi^3$-theory. The main
result is a completely general technique to directly carry out the sum over the
$(n-3)!$ multivariate residues evaluated at the simultaneous solutions of the
scattering equations without solving them explicitly. Our approach is
essentially based on global duality of residues and the Bezoutian matrix. The
validity of the method has been verified through several examples with 
$n\leq 10$ particles. Another salient aspect is that rationality of all final
results is automatically manifest.

This paper suggests several interesting directions for future research on
scattering equations and the CHY formalism. First of all, it is worthwhile to
compare more thoroughly with other recent papers
\cite{Cachazo:2015nwa,Baadsgaard:2015voa,Baadsgaard:2015hia,Huang:2015yka} which
address the same problem. The very clean and symmetric form of polynomial
scattering equations \eqref{EQ_DOLAN_GODDARD} is an immediate invitation to
further systematic studies using algebraic geometry. We believe that a much
deeper understanding of the scattering equations and the CHY formalism may be
gained from a recursive construction. For instance, is it possible to perform
the required polynomial inversions by induction? Moreover, it is intriguing to
investigate the physical meaning of Bezoutian matrices from scattering
equations. We expect that the procedure presented here may be generalized to
loop level in the near future.

\acknowledgments
We thank C. Baadsgaard, E. Bjerrum-Bohr, J. Broedel, F. Cachazo, P.H. Damgaard,
L. Dolan, B. Feng and P. Goddard for discussions. The authors would like to
express a special thanks to the Mainz Institute for Theoretical Physics. M.S. is
grateful to the Niels Bohr International Academy, where a part of this project
was carried out. M.S. acknowledges generous support from both a DFF-Individual
Postdoc grant and a Sapere Aude: DFF-Research Talent grant awarded by the Danish
Council for Independent Research under contract No. DFF-4181-00563. Y.Z. is
supported by the Ambizione grant (PZ00P2\_161341) from Swiss National
Foundation. The research in this paper has also received funding from the
European Research Council under the European Union’s Seventh Framework Programme
(FP/2007- 2013) / ERC Grant Agreement No. 615203, and been partially supported
by the Swiss National Science Foundation through the NCCR SwissMAP.


\begin{thebibliography}{99}
\bibitem{Cachazo:2013gna} 
  F.~Cachazo, S.~He and E.~Y.~Yuan,
  Phys.\ Rev.\ D {\bf 90}, no. 6, 065001 (2014)
  [arXiv:1306.6575 [hep-th]].

\bibitem{Cachazo:2013hca} 
  F.~Cachazo, S.~He and E.~Y.~Yuan,
  Phys.\ Rev.\ Lett.\  {\bf 113}, no. 17, 171601 (2014)
  [arXiv:1307.2199 [hep-th]].

\bibitem{Cachazo:2013iea} 
  F.~Cachazo, S.~He and E.~Y.~Yuan,
  JHEP {\bf 1407}, 033 (2014)
  [arXiv:1309.0885 [hep-th]].

\bibitem{Cachazo:2014nsa} 
  F.~Cachazo, S.~He and E.~Y.~Yuan,
  JHEP {\bf 1501}, 121 (2015)
  [arXiv:1409.8256 [hep-th]].

\bibitem{Cachazo:2014xea} 
  F.~Cachazo, S.~He and E.~Y.~Yuan,
  JHEP {\bf 1507}, 149 (2015)
  [arXiv:1412.3479 [hep-th]].

\bibitem{Weinzierl:2014vwa} 
  S.~Weinzierl,
  JHEP {\bf 1404}, 092 (2014)
  [arXiv:1402.2516 [hep-th]].

\bibitem{Britto:2004ap} 
  R.~Britto, F.~Cachazo and B.~Feng,
  Nucl.\ Phys.\ B {\bf 715}, 499 (2005)
  [hep-th/0412308].

\bibitem{Britto:2005fq} 
  R.~Britto, F.~Cachazo, B.~Feng and E.~Witten,
  Phys.\ Rev.\ Lett.\  {\bf 94}, 181602 (2005)
  [hep-th/0501052].

\bibitem{Dolan:2013isa} 
  L.~Dolan and P.~Goddard,
  JHEP {\bf 1405}, 010 (2014)
  [arXiv:1311.5200 [hep-th]].

\bibitem{Weinzierl:2014ava} 
  S.~Weinzierl,
  JHEP {\bf 1503}, 141 (2015)
  [arXiv:1412.5993 [hep-th]].

\bibitem{Dolan:2014ega} 
  L.~Dolan and P.~Goddard,
  JHEP {\bf 1407}, 029 (2014)
  [arXiv:1402.7374 [hep-th]].

\bibitem{Cachazo:2015nwa} 
  F.~Cachazo and H.~Gomez,
  JHEP {\bf 1604}, 108 (2016)
  doi:10.1007/JHEP04(2016)108
  [arXiv:1505.03571 [hep-th]].

\bibitem{Baadsgaard:2015voa} 
  C.~Baadsgaard, N.~E.~J.~Bjerrum-Bohr, J.~L.~Bourjaily and P.~H.~Damgaard,
  JHEP {\bf 1509}, 129 (2015)
  doi:10.1007/JHEP09(2015)129
  [arXiv:1506.06137 [hep-th]].

\bibitem{Baadsgaard:2015ifa} 
  C.~Baadsgaard, N.~E.~J.~Bjerrum-Bohr, J.~L.~Bourjaily and P.~H.~Damgaard,
  JHEP {\bf 1509}, 136 (2015)
  doi:10.1007/JHEP09(2015)136
  [arXiv:1507.00997 [hep-th]].

\bibitem{Mason:2013sva} 
  L.~Mason and D.~Skinner,
  JHEP {\bf 1407}, 048 (2014)
  [arXiv:1311.2564 [hep-th]].

\bibitem{Berkovits:2013xba} 
  N.~Berkovits,
  JHEP {\bf 1403}, 017 (2014)
  [arXiv:1311.4156 [hep-th]].

\bibitem{Gomez:2013wza} 
  H.~Gomez and E.~Y.~Yuan,
  JHEP {\bf 1404}, 046 (2014)
  [arXiv:1312.5485 [hep-th]].

\bibitem{Ohmori:2015sha} 
  K.~Ohmori,
  JHEP {\bf 1506}, 075 (2015)
  [arXiv:1504.02675 [hep-th]].

\bibitem{Casali:2015vta} 
  E.~Casali, Y.~Geyer, L.~Mason, R.~Monteiro and K.~A.~Roehrig,
  JHEP {\bf 1511}, 038 (2015)
  doi:10.1007/JHEP11(2015)038
  [arXiv:1506.08771 [hep-th]].

\bibitem{Geyer:2014fka} 
  Y.~Geyer, A.~E.~Lipstein and L.~J.~Mason,
  Phys.\ Rev.\ Lett.\  {\bf 113}, no. 8, 081602 (2014)
  [arXiv:1404.6219 [hep-th]].

\bibitem{Lipstein:2015rxa} 
  A.~E.~Lipstein,
  JHEP {\bf 1506}, 166 (2015)
  [arXiv:1504.01364 [hep-th]].

\bibitem{Lipstein:2015vxa} 
  A.~Lipstein and V.~Schomerus,
  arXiv:1507.02936 [hep-th].

\bibitem{Bjerrum-Bohr:2014qwa} 
  N.~E.~J.~Bjerrum-Bohr, P.~H.~Damgaard, P.~Tourkine and P.~Vanhove,
  Phys.\ Rev.\ D {\bf 90}, no. 10, 106002 (2014)
  [arXiv:1403.4553 [hep-th]].

\bibitem{Geyer:2015bja} 
  Y.~Geyer, L.~Mason, R.~Monteiro and P.~Tourkine,
  Phys.\ Rev.\ Lett.\  {\bf 115}, no. 12, 121603 (2015)
  doi:10.1103/PhysRevLett.115.121603
  [arXiv:1507.00321 [hep-th]].

\bibitem{Adamo:2013tsa} 
  T.~Adamo, E.~Casali and D.~Skinner,
  JHEP {\bf 1404}, 104 (2014)
  [arXiv:1312.3828 [hep-th]].

\bibitem{Baadsgaard:2015hia} 
  C.~Baadsgaard, N.~E.~J.~Bjerrum-Bohr, J.~L.~Bourjaily, P.~H.~Damgaard and B.~Feng,
  JHEP {\bf 1511}, 080 (2015)
  doi:10.1007/JHEP11(2015)080
  [arXiv:1508.03627 [hep-th]].

\bibitem{Huang:2015yka} 
  R.~Huang, J.~Rao, B.~Feng and Y.~H.~He,
  JHEP {\bf 1512}, 056 (2015)
  doi:10.1007/JHEP12(2015)056
  [arXiv:1509.04483 [hep-th]].

\bibitem{Kalousios:2015fya} 
  C.~Kalousios,
  JHEP {\bf 1505}, 054 (2015)
  [arXiv:1502.07711 [hep-th]].

\bibitem{Cardona:2015eba} 
  C.~Cardona and C.~Kalousios,
  JHEP {\bf 1601}, 178 (2016)
  doi:10.1007/JHEP01(2016)178
  [arXiv:1509.08908 [hep-th]].

\bibitem{Sogaard:2013fpa} 
  M.~Søgaard and Y.~Zhang,
  JHEP {\bf 1312}, 008 (2013)
  [arXiv:1310.6006 [hep-th]].

\bibitem{Sogaard:2014ila} 
  M.~Sogaard and Y.~Zhang,
  JHEP {\bf 1407}, 112 (2014)
  [arXiv:1403.2463 [hep-th]].

\bibitem{Sogaard:2014oka} 
  M.~Sogaard and Y.~Zhang,
  JHEP {\bf 1412}, 006 (2014)
  [arXiv:1406.5044 [hep-th]].

\bibitem{Johansson:2015ava} 
  H.~Johansson, D.~A.~Kosower, K.~J.~Larsen and M.~Søgaard,
  Phys.\ Rev.\ D {\bf 92}, no. 2, 025015 (2015)
  [arXiv:1503.06711 [hep-th]].

\bibitem{MR507725}
  P.~Griffiths, J.~Harris, ``Principles of Algebraic Geometry''.
  Wiley-Interscience [John Wiley \& Sons], New York,
  1978.

\bibitem{MR0463157}
  R.~Hartshorne, ``Algebraic Geometry". Springer-Verlag, New York,
  1977. Graduate Texts in Mathematics, No. 52.

\bibitem{MR2161985}
  E.~Cattani and A.~Dickenstein. ``Introduction to residues and
  resultants: Solving polynomial equations''. Springer Berlin
  Heidelberg, 2005.

\bibitem{M2}
  D.~R. Grayson and M.~E. Stillman, ``Macaulay2, a software system for research
  in algebraic geometry''. Available at
  http://www.math.uiuc.edu/Macaulay2/.

\bibitem{F4}
  Jean-Charles Faug\`ere, 
  ``A new efficient algorithm for computing Gr{\"o}bner bases (F4)'', 
  Journal of Pure and Applied Algebra, Volume 139, Issues 1–3, Pages 61-88, June 1999.

\bibitem{F5}
  Jean-Charles Faug\`ere,. ``A new efficient algorithm for computing
  Gr{\"o}bner bases without reduction to zero (F5)'' Proceedings of the 2002
  international symposium on Symbolic and algebraic computation (ISSAC)
  (ACM Press): 75–83, July 2002

\bibitem{FGb}
  Jean-Charles Faug\`ere.  FGb: ``A Library for Computing Gr{\"o}bner Bases''.               
  In Komei Fukuda, Joris Hoeven, Michael Joswig, and Nobuki Takayama,               
  editors, Mathematical Software  ICMS 2010, 
  volume 6327 of  Lecture Notes in Computer Science, 
  pages 84-87, Berlin, Heidelberg,               
  September 2010. Springer.

\bibitem{Farey}
  Peter Kornerup and R. T. Gregory, ``Mapping integers and hensel codes
  onto Farey fractions'', BIT Numerical Mathematics, page 9-20, 23, 1983, 
  Kluwer Academic Publishers.

\bibitem{MathematicaM2}
  The package can be downloaded from https://bitbucket.org/yzhphy/mathematicam2.
\end{thebibliography}
\end{document}